\documentclass[12pt,fleqn]{article}
\usepackage{amssymb}
\usepackage{graphicx}
\usepackage{amsmath}
\usepackage{amsfonts}
\usepackage[latin1]{inputenc}

\begin{document}

\begin{centering}
{\leftskip=2in \rightskip=2in
{\large \bf Phenomenology of Doubly Special Relativity}}\\
\bigskip
\bigskip
\bigskip
\medskip
{\small {\bf Giovanni AMELINO-CAMELIA}$^a$, {\bf Jerzy KOWALSKI-GLIKMAN}$^b$,\\
{\bf Gianluca MANDANICI}$^a$
and {\bf Andrea PROCACCINI}$^a$}\\

\bigskip
$^a${\it Dipart.~Fisica Univ.~La Sapienza and Sez.~Roma1 INFN}\\
{\it P.le Moro 2, I-00185 Roma, Italy}\\
$^b${\it Institute for Theoretical Physics,
University of Wroc{\l}aw}\\
{\it pl. Maxa Borna 9, 50--204 Wroc{\l}aw, Poland}

\end{centering}

\vspace{0.7cm}

\begin{center}
\textbf{ABSTRACT}
\end{center}

\baselineskip 11pt plus .5pt minus .5pt

{\leftskip=0.6in \rightskip=0.6in {\footnotesize
Investigations of the possibility that some novel ``quantum" properties
of spacetime might induce a Planck-scale
modification of the energy/momentum dispersion relation
focused at first on scenarios with
Planck-scale violations of Lorentz symmetry, with an associated
reduced $n$-parameter ($n < 6$) rotation-boost symmetry group.
More recently several studies have considered the possibility
of a ``doubly special relativity", in which the modification
of the dispersion relation emerges from a framework with
both the Planck scale and the speed-of-light scale as characteristic
scales of a 6-parameter group of rotation-boost symmetry transformations
(a deformation of the Lorentz transformations).
For the schemes with broken Lorentz symmetry at the Planck scale
there is a large literature on the derivation of experimental limits.
We provide here a corresponding analysis for the doubly-special-relativity
framework. We find that the analyses of
photon stability, synchrotron radiation, and
threshold conditions for particle production in collision processes,
the three contexts which are considered as most promising for
constraining the broken-Lorentz-symmetry scenario,
cannot provide significant constraints on doubly-special-relativity
parameter space.
However, certain types of analyses of gamma-ray bursts
are sensitive to the symmetry deformation.
A key element of our study is an observation that removes a possible
sign ambiguity for the doubly-special-relativity framework.
This result also allows us to characterize more sharply
the differences between the doubly-special-relativity framework and
the framework of $\kappa$-Poincar\'{e} Hopf algebras, two frameworks which
are often confused with each other in the literature.
}}

\vspace{24pt}
\newpage

\baselineskip 12pt plus .5pt minus .5pt
\pagenumbering{arabic}

\setcounter{footnote}{0} \renewcommand{\thefootnote}{\alph{footnote}}

\pagestyle{plain}

\section{Introduction}

Because of its central role in modern physics,
Lorentz symmetry has been investigated in great detail.
The analysis of test theories that could be used as a
measure of our level of experimental
verification of Lorentz symmetry, and could describe violations of
Lorentz symmetry, was already rather mature in the 1940s~\cite{rob}.
The level of interest in
this subject has grown gradually over the last three
decades (see, {\it e.g.}, Refs.~\cite{mans,romansean,gonza,kostcoll,colgla}),
and in particular in these past four or five years
a large number of studies
(see, {\it e.g.},
Refs.~\cite{grbgac,gampul,billetal,polonpap,kifu,mexweave,ita,aus,gactp,jaco,qgpSS,qgpDA,nickcpt}
and references therein)
has explored the possibility
that Lorentz symmetry might be broken by
Planck-scale physics (``quantum gravity").

It is plausible that the unification of
general relativity and quantum mechanics could involve some sort
of spacetime quantization, such as spacetime noncommutativity or
spacetime discreteness, and this can indeed justify some interest in
the fate of Lorentz symmetry at the Planck scale.
If spacetime is fundamentally discrete or noncommutative then
this should in particular apply to
the spacetimes that low-energy probes perceive as Minkowski (flat
classical and continuous), and the recent
quantum-gravity literature provides support
for the expectation that, in the flat-spacetime limit,
spacetime discreteness~\cite{gampul,mexweave,thooftdiscrete}
and spacetime noncommutativity~\cite{majrue,kpoinap,susskind,dougnekr}
are likely to require some departures from Lorentz symmetry.

At first interest in Planck-scale departures from Lorentz symmetry
focused~\cite{grbgac,gampul,billetal,polonpap,kifu,mexweave,ita,aus,jaco,gactp}
on some mechanisms for breaking Lorentz symmetry through
quantum-gravity effects.
These scenarios are characterized by ``symmetry loss":
only a (possibly empty) subset of
the 6-parameter family of transformations that constitute the
Lorentz group would survive as symmetries in the Planck regime.
As proposed in Ref.~\cite{gacdsr}, and verified in a variety of
schemes in Refs.~\cite{gacdsr,dsrnext,leedsr,dsrothers},
it is also possible that Planck scale effects would induce departures
from ordinary Lorentz symmetry which do not however
involve a loss of symmetry.
In this so-called ``doubly special relativity" (``DSR") scenario
one contemplates the possibility that by gaining access
to data involving particles of higher energies we might discover
that in Nature the laws of transformation between inertial observers
are characterized by an invariant Planckian energy/momentum/length
scale, just like nearly a century ago, as data on higher velocities
became available it was established that the speed-of-light scale
is a characteristic of the laws of transformation between inertial
observers.
In Galilei-Newton relativity there is a 6-parameter family
of rotation/boost symmetry transformations
between inertial observers, with laws of transformation
that are scale independent (in the sense that
they do not involve any characteristic scale).
In special relativity one still has a 6-parameter family
of rotation/boost (Lorentz) symmetry transformations, but the laws
of transformation are characterized by an invariant velocity
scale, the speed-of-light scale.
In doubly-special relativity once again one
has a 6-parameter family
of rotation/boost (Lorentz) symmetry transformations, but the laws
of transformation are characterized by two invariant scales,
the speed-of-light scale and the Planck scale.
The doubly-special-relativity rotation/boost transformations
are a Planck-scale ``deformation"
of the special-relativity transformations,
which in turn are a $c$-scale ``deformation" of the Galilei transformations.
In the low-energy limit a DSR scheme turns
into special relativity, just like in the low-velocity limit
special relativity turns into Galilei-Newton relativity.

We focus here on this possibility that at the Planck scale Lorentz symmetry
be deformed (rather than broken).
Specifically we focus on certain types of observations which could allow
to test some of the predictions of the most popular DSR schemes.
For the case of Planck-scale-broken Lorentz symmetry there is already
a rather wide literature on experimental
tests~\cite{grbgac,gampul,billetal,polonpap,kifu,ita,aus,gactp,newlimit,jaconature,stecksync,emnsync}.
For the DSR case, with Planck-scale-deformed Lorentz symmetry,
a corresponding analysis is still missing, and we intend to fill this gap here.

In the next section we start by introducing the most studied model
with Planck-scale-broken Lorentz symmetry
and the most studied model
with Planck-scale-deformed Lorentz symmetry, and we show that these
two models can be very naturally compared since they predict the
same leading-order modification of the dispersion
relation\footnote{While sometimes,
especially when commenting on the logical structure of the DSR framework,
it is convenient for us to indicate explicitly the speed-of-light scale $c$,
in most equations we adopt conventions such that $c = \hbar =1$.}
\begin{equation}
0 \simeq E^2 - \vec{p}^2 - m^2 - \eta \frac{E}{E_p} \vec{p}^2
~,
\label{displead}
\end{equation}
where $\eta$ is a dimensionless coefficient, $E_p$ is the
Planck energy scale ($E_p \simeq 10^{28}eV$), and we
are considering the possibility that such a dispersion relation
would describe the low-energy limit $E \ll E_p$
(we only included a term linear in $E_p^{-1}$).

A key objective of the type of phenomenology that is here of interest
is the one of setting limits on the parameter $\eta$, which cannot be
much smaller than 1 if the conventional quantum-gravity intuition
is to be realized (for $\eta \ll 1$ the scale the characterizes the
onset of the new effects, $E_p /\eta$, would be much bigger than the
Planck scale). In Section~3 we observe that even without resorting
to data one can investigate from a theory perspective the issue
of the ``sign of $\eta$". In particular, we argue that in the
scenario with Lorentz symmetry broken at the Planck scale it is
natural to expect $\eta < 0$, although one cannot completely rule out
the case of positive $\eta$. We also show that in the DSR scenario,
with deformed Lorentz symmetry, $\eta$ must be positive ($\eta > 0$)
in order to have a genuine 6-parameter symmetry group of rotation/boost
transformations, as required by the DSR principles.
The ``sign of $\eta$" had been  considered~\cite{gacdsr,leedsr}
as a key ambiguity for the DSR framework, and we show that this potential
ambiguity can be completely removed.
This proves to be a key asset for the phenomenological analysis
that follows, and also allows us to characterize more sharply
the differences between the doubly-special-relativity framework and
the framework of $\kappa$-Poincar\'{e} Hopf algebras, two frameworks which
are often confused with each other in the literature.

In Section~4 we briefly review the analysis of photon stability
in the scenario with
Lorentz symmetry broken at the Planck scale, and we provide
a corresponding analysis for the DSR scenario.
Although the two scenarios we consider adopt the same leading-order
modification of the dispersion relation, in the broken-symmetry
scenario, for $\eta > 0$, one finds that high-energy photons
can decay into an electron positron pair, whereas
in the DSR scenario the process $\gamma \rightarrow e^+ e^-$
is forbidden.
Some observations in astrophysics allow to establish that the
photon is stable enough to exclude the possibility $\eta > 0$
for the case with broken Lorentz symmetry (therefore in the
broken-symmetry case  $\eta > 0$, which was already disfavored
conceptually on the basis of the points raised in Section~3,
is also ruled out by data).

In Section~5 we consider synchrotron radiation. We review a preliminary
analysis~\cite{jaconature} which has been used to argue that,
for $\eta < 0$, a scheme in which Lorentz symmetry is broken at the Planck
scale should affect significantly the analysis of certain astrophysical
contexts involving synchrotron radiation.
On the basis of our
result of Section~3, showing that $\eta < 0$ is not admissible for
the DSR scenario, we find that instead
synchrotron radiation is not significantly affected in DSR.

In Section~6 we briefly review the analysis of ``threshold anomalies"
in the case of broken Lorentz symmetry, and we also analyze (on the
basis of the corresponding results of Ref.~\cite{gacdsr})
the possibility of threshold anomalies in DSR.
One speaks of a threshold anomaly~\cite{gactp,nguhecr}
when the Planck-scale effects induce a significant modification
of the threshold conditions for particle production in collision processes.
In particular, there has been strong
interest~\cite{kifu,ita,gactp,alfarouhecr,gamboauhecr}
in the possibility that Planck-scale effects might modify the estimate
of the GZK~\cite{gzk} threshold, {\it i.e.} the estimate
of the minimum value of energy needed for a cosmic-ray proton to
interact with a CMBR photon, leading to photopion production.
In the case in which Lorentz symmetry is broken at the Planck scale
one indeed finds~\cite{kifu,ita,gactp,alfarouhecr}
that, for $\eta < 0$,
the GZK threshold is significantly modified, and forthcoming cosmic-ray
observatories could test this effect.
Also in the corresponding DSR analysis there is a modification
of the threshold condition, but it is extremely small, negligible
even for ultra-high-energy cosmic rays.

In Section~7 we consider
certain types of time-of-travel analyses for astrophysical signals
with rich time-versus-energy structure.
In presence of a deformation of the dispersion relation one expects
a (small, Planck-scale suppressed) energy-dependence of the speed
of photons. In the case of a burst of photons of different energies
all emitted  at the same (within a certain accuracy) time from a point
far away from a Earth observatory, one then expects a correlation between time
of arrival on Earth and energy.
The analysis in this context proceeds basically in the same way
independently of whether Lorentz symmetry is broken or deformed.
However, in Sections~3 and 4 we established that the broken-symmetry
scenario requires negative $\eta$ whereas the DSR scenario requires
positive $\eta$. This ``sign difference" opens the way for
studies able to discriminate between the broken-symmetry and the DSR
scenarios using the next generation of gamma-ray observatories
(such as the GLAST space telescope, which is already planning~\cite{glast}
this type of time-of-arrival studies motivated by Planck-scale physics).

In the last section (Section~8) we summarize our key results and
comment on the outlook of this research programme.

\section{Broken versus deformed Lorentz symmetry}



\subsection{A scenario for Planck-scale-broken Lorentz symmetry}
There is growing
interest~\cite{grbgac,gampul,billetal,polonpap,kifu,mexweave,ita,aus,gactp,jaco,qgpSS,qgpDA,nickcpt}
in the possibility that some novel quantum properties of
spacetime may have important implications for the analysis of
Lorentz transformations. Some approaches to the quantum-gravity
problem attribute to the Planck scale $E_p$
the status of an intrinsic characteristic of space-time
structure.
For example $E_p$ can have a role in spacetime
discretization or in the commutation relations between spacetime
observables. In discretized versions or non-commutative versions of
Minkowski space-time it is not uncommon to find departures
from ordinary Lorentz symmetry (see, {\it e.g.}, the pedagogical
discussions in Ref.\cite{thooftdiscrete} for the case of
discretization and in Refs.\cite{majrue,kpoinap,susskind,alessfranc} for
the case of non-commutativity). The action of ordinary (classical)
boosts on discretization length scales (or non-commutativity
length scales) can be such that different inertial
observers would attribute different values to these lengths
scales, just as one would expect from the mechanism of
FitzGerald-Lorentz contraction.

Models based on an approximate Lorentz symmetry, with
Planck-scale-dependent departures from exact Lorentz symmetry,
have been recently considered in most quantum-gravity research
lines, including models based on ``spacetime foam"
pictures~\cite{grbgac,garayPRL}, ``loop quantum gravity"
models~\cite{gampul}, certain ``string theory"
scenarios~\cite{susskind,bertoNC}, and ``noncommutative
geometry~\cite{susskind,gacdsr,dsrnext}.

The most studied model of Planck-scale departures from Lorentz symmetry
is the one which evolved primarily through the studies reported in
Refs.~\cite{grbgac,kifu,ita,gactp}. This is a kinematic in which
the Planck-scale $E_p$ enters the energy/momentum dispersion
relation as in (\ref{displead}), which we report here for convenience
\begin{equation}
0 = f(E,\vec{p}^2,m;E_p)
\simeq E^2 - \vec{p}^2 - m^2 - \eta \frac{E}{E_p} \vec{p}^2
~,
\label{displeadbis}
\end{equation}
while the laws of energy-momentum conservation remain
unaffected by the Planck scale, and therefore for example
in a process $a + b \rightarrow c + d$
\begin{eqnarray}
E_a + E_b = E_c + E_d \nonumber\\
\vec{p}_a + \vec{p}_b = \vec{p}_c + \vec{p}_d
~.
\label{epconsnormal}
\end{eqnarray}
In different quantum-gravity models with Planck-scale departures
from Lorentz symmetry the exact dispersion relation (the exact form
of the function $f(E,\vec{p}^2,m;E_p)$) will in general be different,
but
it has been argued that for a variety of quantum-gravity models one should
find that for $E \ll E_p$ (small
energies) $f \simeq E^2 - \vec{p}^2 - m^2 - \eta E \vec{p}^2/E_p$,
as in (\ref{displeadbis}).
The fact that in this popular scenario one combines the
modified dispersion relation (\ref{displeadbis}) with the
unomdified law of energy-momentum conservation (\ref{epconsnormal})
is an explicit indication of the assumed breaking of Lorentz
symmetry. There is in fact no 6-parameter family of rotation/boost-like
symmetry transformation which are compatible with
both requirements (\ref{displeadbis}) and (\ref{epconsnormal}).
(We elaborate more on this point in the next subsection).

The fact that the literature has focused primarily on this scenario is
mostly due to its simplicity, which makes it a natural first step in a
phenomenology of Planck-scale departures from Lorentz symmetry.
However, this scenario is also more or less directly connected
with various quantum-gravity proposals.
In Loop Quantum Gravity preliminary results~\cite{gampul,leeDispRel}
provide support for the dispersion relation (\ref{displeadbis}).
Proposals based on noncommutative spacetimes inevitably
lead to departures from ordinary Lorentz
symmetry and there are various examples~\cite{kpoinap}, in which one
finds (\ref{displead}).
In String Theory it appears that the modification of the dispersion
relation is not automatic but emerges in presence of
certain natural background fields~\cite{susskind,bertoNC}, and for some
background configurations a dispersion relation of the
type (\ref{displead}) is encountered~\cite{bertoNC}.
The role that phenomenological analyses
of this scenario (\ref{displead}), (\ref{epconsnormal})
could have in the overall development of
quantum-gravity research has been stressed in most recent reviews
by experts of the field (see, {\it e.g.},
Refs.~\cite{carlorev,carliprev,leerev}).

At an intuive level this scenario can be seen in analogy with the
emergence of deformed
dispersion relations in the description of collective modes
(such as phonons) in certain materials, whose propagation satisfies a
relativistic dispersion relation only up to corrections governed by the
scale of atomic structure of the material\footnote{Similar
considerations apply to the decription of the propagation of
light in water.}. Intuitively one can indeed attempt to introduce in quantum
gravity the concept of ``spacetime foam", and spacetime foam
might affect particle propagation in a way that to some extent can be viewed
in analogy with the way that the presence of other media affects particle
propagation.

\subsection{A corresponding doubly-special-relativity scenario}
The fact that the quantum-gravity scenario discussed in the preceding
subsection ``breaks" Lorentz symmetry, {\it i.e.} requires a preferred
class of inertial observers, is not
necessarily a disappointment. As mentioned, certain quantum-gravity ideas,
notably some perspectives on the spacetime-foam picture,
can provide motivation for exploring this possibility.
Moreover, in the study of particle physics the concept of spontaneous
breaking of a symmetry has proven very useful, and it is conceivable that
quantum gravity would host a similar mechanism for the spontaneous
breaking of Lorentz symmetry at the Planck scale.
For example, in string theory, the most popular
approach to the quantum-gravity problem,
mechanisms for the spontaneous breaking of Lorentz symmetry
have been investigated
(see, {\it e.g.}, Ref.~\cite{kostcoll,dougnekr} and references therein).

However, the existence of a preferred class of inertial observers
is anyway not simple conceptually.
At present we can only make wild guesses about this
preferred class of observers.
Many studies attempt to
identify this preferred class
of inertial observers with a natural class of observers for
the CMBR, but this conjecture (although it cannot be excluded a priori)
appears to lack any justification, since it would be rather surprising to
find a connection between the physics responsible for the CMBR
and the Planck-scale realm.

Even if one is not troubled by the possibility
of a preferred class of inertial observers,
it is natural to wonder~\cite{gacdsr} whether
there are other alternatives in addition to the cases in which
Lorentz symmetry is exactly preserved at the Planck scale and the
case in which Lorentz symmetry is broken, with associated
emergence of a preferred class of inertial observers.
The so-called ``doubly-special relativity" (``DSR")
theories were proposed~\cite{gacdsr} as a scenario
in which {\bf (i)} the Planck energy $E_p$ takes the role
of relativistic invariant, in the same sense that the speed-of-light
scale ``$c$" is a relativistic invariant, and {\bf (ii)} ordinary Lorentz
symmetry is not exactly preserved at the Planck scale, but
there is no special class of inertial observers.
As mentioned, this DSR proposal
can be described, using a terminology which is popular in
the mathematics community,
as a ``deformation"
of Lorentz symmetry, without any actual ``loss of symmetry".
This means that the Lorentz laws of transformation
between inertial observers are replaced by different laws of
transformation between inertial observers (in the DSR case this
involves the introduction of a second observer-independent scale $E_p$)
but without any loss of equivalence between inertial observers:
all inertial observers remain equivalent (there is no preferred
class of inertial observers), but the laws of transformation
between inertial observers are modified with respect to the
original Lorentz transformation laws.

It is emerging that the DSR framework is needed in order to
understand some aspects of certain noncommutative
spacetimes~\cite{gacdsr,dsrnext},
and possibly also of the ``loop quantum gravity"
approach~\cite{kodadsr}.
Here, since we are focusing on phenomenological issues,
we prefer to stress that the DSR hypothesis
can be viewed as a logical continuation of the path that already connects
Galilei-Newton relativity and special relativity.
In Galilei-Newton relativity there is no observer-independent scale
characteristic of the rotation/boost transformations,
and in fact the dispersion relation is written
as $E=p^2/(2m)$ (whose structure fulfills the requirements
of dimensional analysis without the need for dimensionful
coefficients). As experimental evidence in favour of Maxwell equations
started to grow, the fact that those equations involve a
special velocity scale appeared to require (assuming the Galilei
symmetry group should remain unaffected) the introduction
of a preferred class of inertial observers (the ``ether").
However, in the end we discovered that the Maxwell theory
does not require a preferred class
of inertial observers, but rather
it was wrong to assume that Galileian relativity should apply
in all regimes. In the high-velocity regime it must be replaced
by special relativity.
Special relativity introduces the first observer-independent scale,
the velocity scale $c$, and its dispersion relation
takes the form $E^2 = c^2 p^2 + c^4 m^2$.
As interest in dispersion relations
of the type $0 =E^2 -  c^2 \vec{p}^2 - c^4 m^2 + f(E,\vec{p}^2,m;c,E_p)$
is starting to grow within the quantum-gravity community
(also because of the analysis of noncommutative spacetimes and loop
quantum gravity)
the fact that these dispersion relations involve a special energy
scale, $E_p$, was leading to the assumption that
a preferred class of inertial observers might have to be introduced.
The DSR proposal essentially raises the possibility that the
assumption of a preferred class of inertial observers might once again
be incorrect, and once again we might need to deform the
laws of transformation between inertial observers (as already done in going
from Galilei-Newton relativity to special relativity).
A dispersion relation of the
type $0 =E^2 -  c^2 \vec{p}^2 - c^4 m^2 + f(E,\vec{p}^2;E_p)$
can in fact hold for all inertial observers~\cite{gacdsr,dsrnext}.

In particular, the dispersion relation\footnote{This type
of dispersion relation was considered in some of the first
papers on the DSR proposal~\cite{gacdsr,dsrnext}, and remains
the focus of much of DSR research.
As mentioned, the leading-order form of this dispersion relation had been
previously considered for the broken-Lorentz-symmetry scenario
of Ref.~\cite{grbgac}, and the full exact form
of this dispersion relation had been considered even earlier
in the Hopf-algebra literature~\cite{majrue,kpoinap}.}
\begin{equation}
0= \frac{2}{\lambda^2} \left[\cosh (\lambda E)
- \cosh (\lambda m ) \right]
- \vec{p}^2 e^{\lambda E}
\simeq E^2 - \vec{p}^2 - m^2 - \lambda E \vec{p}^2
\label{dispKpoin}
\end{equation}
can be valid in all inertial frames, at the cost of a $\lambda$-dependent
deformation of the boost transformations~\cite{gacdsr,dsrnext}.
And one easily notices that for $\lambda \equiv \eta /E_p$
the dispersion relation (\ref{dispKpoin}) is consistent with
the dispersion relation (\ref{displead}) considered in the popular
broken-Lorentz-symmetry scenario discussed in the previous
subsection.

The fact that the same dispersion relation can be considered both in
a broken-Lorentz-symmetry scenario and in a deformed-Lorentz-symmetry
scenario raises an interesting challenge from a phenomenological
perspective: how can we distinguish experimentally between
the two scenarios?
We will show that, although (as established
previously~\cite{grbgac,gampul,billetal,polonpap,qgpSS,qgpDA})
the  modification of the dispersion relation is
the most important ingredient
of the strategies for testing Planck-scale departures
from ordinary Lorentz symmetry,
it is possible to distinguish experimentally between a DSR scheme
and a broken-Lorentz-symmetry scheme which adopt the same
Planck-scale modification of the dispersion relation.
Of course, one must exploit the fact other aspects of kinematics
(in addition to the dispersion relation)
must also be Planck-scale modified in a DSR scheme,
for consistency with the request of equivalence among inertial frames
(whereas the same is not true for the
broken-Lorentz-symmetry scenario).

We will stress that, in particular, the unmodified law of energy-momentum
conservation, which is assumed in the broken-Lorentz-symmetry scenario
of the previous subsection, is not compatible with the DSR requirements.
An easy way to show this incompatibility is based on the form of
the dependence of energy-momentum on
the rapidity parameter $\xi$ (the coefficient
of a boost generator $N$ in the exponentiation $e^{\xi N}$ that
implements a finite boost transformation).
For a DSR scenario with dispersion relation (\ref{dispKpoin})
one finds~\cite{gacdsr,dsrnext} that the rapidity/energy-momentum
relation is such that
\begin{equation}\label{xidsr1}
\cosh (\xi) = \frac{e^{\lambda E} - \cosh\left(\lambda m\right)}
  {\sinh\left(\lambda m \right)} ~,~~~
\sinh (\xi) = \frac{\lambda p e^{\lambda E}}
  {\sinh\left(\lambda m\right)}\,\, ,
\end{equation}
where $\xi$ here is the amount of rapidity needed to take a particle
from its rest frame ($E=m$, $p=0$)
to a frame in which its energy is $E$ (and its momentum
is $p(E)$, which is fixed in terms of $E$ by the dispersion relation
and the direction of the boost).
Of course, in the $\lambda \rightarrow 0$ limit these relations
reproduce the corresponding special relativistic relations
\begin{equation}\label{xisre}
\cosh (\xi) = \frac{E}{m} ~,~~~
\sinh (\xi) = \frac{p}{m} \,\, .
\end{equation}
And, as needed for the DSR requirement,
one can easily verify that the $E(\xi)$ and $p(\xi)$ implicitly
defined by (\ref{xidsr1}) satisfy the dispersion relation  (\ref{dispKpoin})
for every value of $\xi$ (the $\lambda$-modified dispersion relation
(\ref{dispKpoin}) holds in every frame connected by the relevant boost).
However, one can also easily verify that it is not possible to
have $E_a(\xi) + E_b(\xi) = E_c(\xi) + E_d(\xi)$,
$\vec{p}_a(\xi) + \vec{p}_b(\xi) = \vec{p}_c(\xi) + \vec{p}_d(\xi)$
(in a process $a+b \rightarrow c+d$) for every $\xi$.
If one was to enforce the unmodified law of energy-momentum conservation
in a framework with boost transformations of type (\ref{xidsr1})
then this unmodified law of energy-momentum conservation could
only hold in one inertial frame (it would be violated
in other frames reacheable by boosting).

This incompatibility (within a DSR framework) between modified
dispersion relation and unmodified energy-momentum conservation
was already noticed in the first papers on DSR~\cite{gacdsr},
leading to the realization that in order to have the equivalence
of all inertial frames the presence of a modification of the
dispersion relation required a corresponding modification
of the law of energy-momentum conservation.
For the specific DSR scheme we are considering (with dispersion
relation (\ref{dispKpoin}) and boost transformations codified
in (\ref{xidsr1})) this needed modification\footnote{Since
we stressed the analogy between the transition
Galilei relativity $\rightarrow$ special relativity
and the conjectured transition
special relativity $\rightarrow$  doubly special relativity,
it is perhaps worth mentioning that this DSR modification of
energy-momentum conservation must be viewed in analogy with
the special-relativistic modification of the law of composition of
velocities.
In Galilei relativity there is no special velocity scale, and inevitably
the law of composition of velocities must take the form $V_0 + V$
(where, in particular, $V$ could be the velocity of a particle
 in a given inertial frame $O$ and  $V_0 + V$ could be the velocity
 of that same particle  in another inertial frame $O'$, if $V_0$ is
 the relative velocity for the frame $O$ and $O'$).
In special relativity there is a special velocity scale $c$ and the law
of composition of velocities becomes nonlinear and $c$-dependent
(it must saturate at the maximum velocity $c$).
In turn in special relativity there is no special energy-momentum scale,
and therefore the law of composition of energy-momentum
(in particular in obtaining the total momentum of a two-particle system)
must be linear, $P_\mu + P'_\mu$.
But then in DSR there is special energy-momentum scale, $1/\lambda$,
and the the law of composition of energy-momentum
becomes nonlinear and $\lambda$-dependent.}
of the law of energy-momentum
conservation has been worked out~\cite{gacdsr,dsrnext}.
The exact form of this law is a rather messy combination of
exponentials, but fortunately in the following sections we will only
the need the leading-$\lambda$-order form of the law
of energy-momentum conservation, which is~\cite{gacdsr,dsrnext}
\begin{equation}
E_a + E_b - \lambda p_a p_b
- E_c -E_d + \lambda p_c p_d = 0
~,
\label{conservnewe}
\end{equation}
\begin{equation}
p_a + p_b - \lambda (E_a p_b + E_b p_a)
- p_c -p_d + \lambda (E_c p_d + E_d p_c) = 0
~.
\label{conservnewp}
\end{equation}
This law of energy-momentum conservation is indeed compatible
with the DSR boost transformations; in fact, using (\ref{xidsr1})
one can easily verify that (to leading order in $\lambda$)
if energy-momentum is conserved in
one inertial frame then it is automatically conserved also
in all other inertial frames.

This observation already provides us a valuable
key for phenomenological analyses. In
the popular broken-Lorentz-symmetry scenario of the previous subsection
one has the modified dispersion relation  (\ref{dispKpoin})
with unmodified law of energy-momentum conservation.
A DSR scheme can adopt the same modified dispersion
relation (\ref{dispKpoin}) but then for consistency with the required
equivalence of inertial frames it must also impose a modified
law of energy-momentum conservation (\ref{conservnewe})-(\ref{conservnewp}).

In the phenomenology discussed in the following Section~4, 5, 6 and 7 this
previous result on the law of energy-momentum conservation in DSR
will play a key role. In addition we will also use
a new result which we obtain in the next Section~3:
we will show that in the DSR framework
one must necessarily assume $\lambda E_p > 0$
({\it i.e.} the dimensionful parameter $\lambda$ that appears
in (\ref{dispKpoin}) and (\ref{xidsr1}) must necessarily be positive).
This result proves very useful since the possibility that,
in the broken-Lorentz-symmetry scenario of the previous subsection
the parameter $\eta$
be positive (and we remind the reader that $\eta \sim \lambda E_p$)
is already excluded experimentally (see Section~4).

\subsection{Other schemes for Planck-scale departures from
Lorentz symmetry}
There are clearly three possibilities for the fate of Lorentz
symmetry at the Planck scale:
 {\bf (I)} Lorentz symmetry, with its dispersion relation and
 its other key features, remains unmodified even at the Planck scale
 {\bf (II)} Lorentz symmetry is broken by Planck-scale effects,
 with associated emergence of a preferred class of inertial observers,
and {\bf (III)} Lorentz symmetry is deformed, in the DSR sense,
preserving the equivalence of inertial observers.

The scenario discussed in Subsection~2.1 is the most studied
broken-Lorentz-symmetry scenario, while the
scenario discussed in Subsection~2.2 is the most studied
DSR (deformed-Lorentz-symmetry) scenario.
But there are other ways to break Lorentz symmetry at the
Planck scale which have been considered in the literature,
and there are other DSR scenarios which have been considered in the
literature.
In particular, both in work on broken Lorentz symmetry and in
work on DSR, there has been some interest in the possibility
that the dispersion relation be modified more softly, {\it i.e.}
that the term linear in $\lambda$ ($\eta /E_p$) might be absent,
leading to a dispersion relation which at low energies takes
the form
\begin{equation}
0 \simeq E^2 - \vec{p}^2 - m^2 + \eta_2 \frac{E^2}{E_p^2} \vec{p}^2
~.
\label{dispQUADRA}
\end{equation}
Although we are choosing to focus on the most studied scenarios,
the ones of Subsections~2.1 and 2.2,
in the following Sections
we shall sometimes briefly comment of the possibility (\ref{dispQUADRA}).

Especially in the DSR framework there is a line of analysis
that can lead to considering~\cite{dsrgzk}
a scenario with a very mild modification of the
dispersion relation at energies low enough that $E \leq (m^j E_p^{n-j})^{1/n}$
(where $n$,$j$ are some integers, while $m$ and $E_p$ denote again
the mass of the particle and the Planck scale respectively)
but a rather significant modification of the dispersion relation
for energies such that $(m^j E_p^{n-j})^{1/n} < E < E_p$.
In this case also the relation between rapidity and energy (in the
sense of (\ref{xidsr1})) would take a form such that it appears
to be nearly exactly special relativistic for $E \leq (m^j E_p^{n-j})^{1/n}$
but is significantly modified for $(m^j E_p^{n-j})^{1/n} < E < E_p$.
An example of this type of rapidity-energy relation
was considered in Ref.~\cite{dsrgzk}:
\begin{equation}\label{dispgood}
\cosh (\xi) = {(E /m)} (2 \pi)^{-E^2 \tanh[m^2 E_p^4/E^6]/(m E_p+E^2)} \,\, .
\end{equation}
We will only briefly comment on the phenomenological
implications of this possibility in Section~6 (in association with
the GZK threshold for ultra-high-energy cosmic rays).

Besides possible alternative forms of the Planck-scale-dependent terms
in the dispersion relation,
there has been also some interest in the literature in ``nonuniversal"
dispersion relations. The scenarios on which we focus (the ones
of Subsections~2.1 and 2.2) are implicitly ``universal", in
the sense that the relation between energy and momentum depends
(once $c$ and $E_p$ are fixed) only on the mass of the particle,
and not on other properties of the particle, such as spin and
electromagnetic charge.
Especially from a broken-Lorentz-symmetry perspective,
some authors (see, {\it e.g.}, Refs.~\cite{jaconature,emnsync})
have considered the possibility of a ``nonuniversal"
Planck-scale modification of the dispersion relation,
{\it i.e.} a dispersion relation of type (\ref{displead})
(or (\ref{dispQUADRA})) in which however the coefficient $\eta$ (or $\eta_2$)
is different for different particles. For example, the value of $\eta$
for photons could be different~\cite{jaco,jaconature,emnsync}
from the value of $\eta$ for electrons.
We will comment briefly on this possible ``nonuniversality"
in the sections that report our phenomenological analysis.

\section{Removing the sign ambiguity}
Of course, the key objective of this phenomenology is to establish
whether or not $\eta \neq 0$ in the broken-Lorentz-symmetry
scenario (or $\lambda \neq 0$ in the DSR scenario).
At the next level of priority clearly it is important to
establish ``the sign of the modification", {\it i.e.} to establish
whether $\eta \leq 0$ or $\eta \geq 0$ (and similarly
for $\lambda E_p \leq 0$ or $\lambda E_p \geq 0$).

As announced earlier, one of the key results of the analysis we are
reporting is that $\eta$ (the parameter of the broken-Lorentz-symmetry
scenario) must be smaller or equal to $0$,
while $\eta_{DSR} \equiv \lambda E_p$
(the parameter of the DSR
scenario) must be greater or equal to $0$.

The condition $\eta \leq 0$ is favoured conceptually, as argued in
the next Subsection (3.1), and one then finds, as we show in Section~4,
that $\eta > 0$ is excluded by data.

As we show in Subsection~3.2,
the condition $\eta_{DSR} \equiv \lambda E_p \geq 0$
is obtained already at the level of the mathematical consistency
of the DSR framework. For $\eta_{DSR} < 0$
it is not possible to enforce
the dispersion relation (\ref{dispKpoin}) in all inertial
frames.

\subsection{A natural sign choice for broken
Lorentz symmetry}
As mentioned, at least at an intuive level, the idea
of Planck-scale broken Lorentz symmetry
is usually viewed in analogy with the
emergence of modified
dispersion relations in the description of
the propagation of particles in certain material media
(as in the case of the propagation of light in water or in
certain crystals).
One can indeed attempt to introduce in quantum
gravity the concept of ``spacetime foam" and spacetime foam
might affect particle propagation in a way that to some extent can be viewed
in analogy with the way that the presence of material media affects particle
propagation.

The sign of $\eta$ in the dispersion relation will fix (making
the natural assumption that the relation $v = dE/dp$ still holds)
whether or not $c$ is still the maximum velocity.
In the analogous situation of dispersion induced by a material medium
one naturally finds that $c$ is still the maximum speed.
The presence of the medium does not change the fact that
the theoretical framework at the fundamental level constrains speeds
to be lower than $c$.
In a broken-Lorentz-symmetry scenario Lorentz symmetry still plays
a role at the fundamental level of the theory, but the presence
of some medium allows to select a preferred frame and makes room
of the possibility that the speed of photons be different from $c$.
But this speed still cannot exceed $c$, in light of the fundamental
constraint imposed by the Lorentz symmetry of the fundamental
theory (before the backgorund is introduced.

This reasoning (however limited, since it is simply based on an
analogy) suggests that in the scenario with broken Lorentz symmetry
at the Planck scale the sign of $\eta$ should negative.
This theoretical expectation is confirmed by data, in the sense
that, as we will discuss in Section~4, consistency with
certain observations in astrophysics requires that $\eta < 0$
in the broken-Lorentz-symmetry scenario of Subsection~2.1.

\subsection{The necessary sign choice
for doubly special relativity}
Just like in the case of broken Lorentz symmetry, for a doubly special
relativity framework the sign choice which
specifies whether or not the relativistic invariant $c$ sets the
maximum value of speed is one of the most important features.
In particular, in the DSR scenario
discussed in Subsection 2.2 (with the dispersion relation (\ref{dispKpoin})
and the dependence of energy-momentum on rapidity governed
by (\ref{xidsr1})) using $v=dE/dp$ one finds
that for negative $\lambda$ the invariant $c$ is the maximum velocity,
which is achieved by massless particles in the low-energy ($E \rightarrow 0$)
limit (and the speed of massless particles decreases with energy).
For positive $\lambda$ the invariant $c$ is still the speed
of massless particles in the low-energy
limit, but the speed of  massless particles increases with energy.

While in the broken-Lorentz-symmetry case it is puzzling to find speeds
higher than $c$ (for the reason discussed in the previous subsection),
in a doubly-special-relativity scenario there is no in-principle
obstruction for speeds higher than $c$. In fact, in doubly special
relativity $c$ is defined operatively as the
speed of massless particles in the low-energy
($E \rightarrow 0$) limit, and there is no {\it a priori} reason
for assuming a description of $c$ as maximum speed.

So far the choice between positive and negative $\lambda$ has been
treated~\cite{gacdsr,dsrnext,leedsr}
as a free choice allowed by the DSR framework, but we intend to show here
that actually only for positive $\lambda$ one obtains a scenario
which is genuinely consistent with the DSR requirements.

This point comes from a simple analysis of the DSR
laws of transformation between inertial observers.
A first indication comes already from the structure of
the equations (\ref{xidsr1}). These equations can be easily
derived by imposing that for every value
of rapidity the DSR energy/momentum dispersion relation
would be satisfied. However, we observe that for negative $\lambda$
the equations (\ref{xidsr1}) correspond to a satisfactory behaviour
only for relatively small rapidity, and for a critical {\underline{finite}}
value of rapidity a divergence of energy in encountered.

In an effort to find the root of this problem, we observe that
the differential equations that govern
the dependence of energy-momentum on rapidity are of a type that
does not necessarily lead to the existence of
global solutions $E(\xi)$, $p(\xi)$,
since the structure of the equations does not
fulfill the standard
Cauchy requirements
for the existence of global solutions $E(\xi)$, $p(\xi)$.
It is sufficient for us to discuss this issue considering a
single boost (along a given direction).
In this case one finds~\cite{gacdsr,dsrnext}
that in the DSR framework
the dependence of energy-momentum
on rapidity is governed by the differential equations
\begin{equation}
p'(\xi) = \frac{\lambda}{2} p^2(\xi)
+ \frac{1-e^{-2\lambda E(\xi)}}{2\lambda}
~, \label{finarfin}
\end{equation}
\begin{equation}
E'(\xi) = p(\xi) ~, \label{finarfinE}
\end{equation}
where we used the standard notation $f'(\xi) \equiv df/d\xi$
(for any function $f$).

For the context we are considering the Cauchy requirements
can be compactly stated
introducing the two-component
function $Y(\xi) \equiv \{ Y_1(\xi), Y_2(\xi) \} \equiv \{E(\xi),p(\xi) \}$,
and using the notation $Y_l'=F_l(Y)$ ($l \in \{ 1,2 \}$)
to denote compactly our system of equations
(\ref{finarfin})-(\ref{finarfinE}):
\begin{itemize}
\item  (i) F must be continuous;
\item (ii) for every $M \in \Re$ and for every  $X \in \Re^2$
and $ Z \in \Re^2$,
such that $|X| \leq M$,$|Z | \leq M$,
there must exist an $L_M \in \Re$,
such that $ \mid F(X)-F(Z) \mid \leq L_M \mid X-Z \mid$;
\item (iii) for every $X \in \Re ^2 $ there must exist $L_1\in \Re$
and $L_2 \in \Re$ such that $\mid F(X)\mid \leq L_1 + L_2 \mid X \mid $.
\end{itemize}
(Of course, with $|W|$ we are
 denoting $\sqrt{W_a^2 + W_b^2}$ for
 every $W \equiv \{ W_a , W_b \} \in \Re ^2 $.)

The requirements (i) and (ii) are easily verified, but a
possible problem for (iii) originates from the nonlinear
structure of our equation (\ref{finarfin}). The corresponding differential
equations of ordinary special relativity ($\lambda \rightarrow 0$ limit)
are linear and automatically verify
the Cauchy ``sublinearity requirement" (iii).
Instead the nonlinearity of the DSR differential equations imposes
a detailed analysis.
Our system of equations (\ref{finarfin})-(\ref{finarfinE})
evidently satisfies the Cauchy requirements for existence and
uniqueness of a {\underline{local}} solution (in a neighborhood
of a given value of $\xi$), but we are not {\it a priori}
assured of the existence of a global solution.

A detailed analysis shows that for positive $\lambda$ there is no
problem: the Cauchy ``sublinearity requirement" (iii) is satisfied
(in spite of the nonlinearity of the equations) and therefore the
existence of global solutions is assured. But for negative $\lambda$
the Cauchy ``sublinearity requirement" is not satisfied.
The interested reader can straightforwardly (but somewhat tediously)
verify that indeed for positive $\lambda$ one can find two real
numbers $L_1$ and $L_2$ with he property required in (iii).
Instead for negative $\lambda$ the requirement (iii) is not
satisfied, for any pair of real numbers $L_1$ and $L_2$.

It is for us here sufficient to discuss
a simplified proof, restricting
our interest to the case relevant for on-shell particles,
in which energy and momentum satisfy the
dispersion relation (\ref{dispKpoin}).
Imposing the dispersion relation one can of course reduce
our system of two differential equations to a single
differential equation:
\begin{equation}
E'(\xi) = \frac{1}{|\lambda |} \sqrt{1- 2 \cosh (\lambda m)
e^{- \lambda E(\xi)}
+ e^{- 2 \lambda E(\xi)}}
~. \label{jocSINGLE}
\end{equation}
Here the Cauchy ``sublinearity requirement" asks us to
find a pair of real numbers $L_1$ and $L_2$ such
that $E'(\xi) \leq L_1 + L_2 E(\xi) $
for every $E(\xi)$.
Indeed for positive $\lambda$ (where the exponentials
in (\ref{jocSINGLE}) are of the type $e^{- |\lambda| E}$,
and $e^{- |\lambda| E} \leq 1$) one can find such pairs of real numbers.
For example, the choice $L_1 = 1/ \lambda$ and $L_2 =0$
is acceptable.
Instead in the case of negative $\lambda$ one finds that
for any given pair of real numbers $L_1$ and $L_2$
there is always a value of $E$ such that,
according to (\ref{jocSINGLE}), $E'(\xi) > L_1 + L_2 E(\xi) $.
This is due to the fact that for negative $\lambda$ the exponentials
in (\ref{jocSINGLE}) are of the type $e^{|\lambda| E}$, and diverge
exponentially for large $E$.

This leads us, as anticipated, to the conclusion that,
in the framework\footnote{In this subsection we provided the
solution for the ``sign ambiguity"
of the DSR scenario of Subsection 2.2.
We are finding that an analogous result also holds in
DSR scenarios that adopt a dispersion relation that is different
from the one of Subsection 2.2. In particular, the dispersion
relation $[m^2/(1- {\tilde{\lambda}} m)^2] = [E^2 - p^2/(1
- {\tilde{\lambda}} m)^2]$
is adopted in a DSR scenario considered in Ref.~\cite{leedsr},
and the issue of the choice of sign
of ${\tilde{\lambda}}$ was indeed raised. Also in that case our line
of analysis allows to fix the sign of the deformation parameter
(the careful reader will easily find that a genuine DSR scenario
is obtained for positive ${\tilde{\lambda}}$, while the possibility
of a negative ${\tilde{\lambda}}$ is not acceptable).}
discussed in Subsection 2.2, for
negative $\lambda$ one does not genuinely obtain a DSR scenario,
since the absence of global solutions for our system of differential
equations would lead to the paradox
that a particle with well-defined energy-momentum
for certain inertial observers, would not have a well-defined
momentum for some other observers.
For positive $\lambda$ there is no such problem and
all the DSR requirements are satisfied.

%
%

\subsection{An illustrative example of $\kappa$-Poincar\'{e}
algebra which is not admissible in DSR}
While the objectives of the analysis we are reporting are primarily
phenomenological, it seems appropriate to devote this subsection
to the discussion of some conceptual (rather than phenomenological)
implications of the result obtained in the previous subsection.

As clarified earlier, the proposal~\cite{gacdsr} of
doubly special relativity is the idea that the quantum-gravity problem
might lead us to the introduction of a second relativistic invariant,
a small-length/large-energy scale, possibly given by the Planck scale.
And the DSR (rotation-)boost transformations should be characterized
by two scales, $\lambda$ ($1 / \lambda$) and $c$, in the same sense that $c$
already characterizes the (rotation-)boost transformations of special
relativity.
It is interesting to ask which types of mathematical formalisms
could play a role in the construction of the physics idea
of a doubly special relativity.
Einstein's special relativity can rely
on the mathematics of Lorentz and Poincar\'{e}.
For doubly special relativity such a clear conclusion has not yet
been reached, but it appears that to some extent the mathematics
of ``$\kappa$-Poincar\'{e} Hopf
algebras"~\cite{kpoinap,lukie}
can play a role in the description of some aspects of
a DSR theory, at least in the cases (as the one we are here considering)
in which the second relativistic invariant appears in a deformation
of the dispersion relation.

This observation has led to some confusion: sometimes the concept of
a DSR physical theory and the concept of $\kappa$-Poincar\'{e} Hopf
algebra are treated interchangeably, as if they were the same concept.
This would anyway be inappropriate in the same sense that the physical
proposal of Einstein's
special relativity cannot be strictly identified with the mathematics
of Lorentz and Poincar\'{e}. And it is even more inappropriate
since in the DSR/$\kappa$-Poincar\'{e}
context one can easily verify that there some aspects of $\kappa$-Poincar\'{e}
mathematics that are not compatible with the DSR physical principles.
Some of these incompatibilities have already been emphasized in
the literature. In particular, it is well known~\cite{dsrnext,kpoinap}
that the laws of conservation of energy-momentum for multiparticle
processes adopted
in the $\kappa$-Poincar\'{e} literature (see Ref.~\cite{lukieNEWdsr}
and references therein) are incompatible~\cite{gacperspe}
with the DSR requirements.
The result obtained in the previous subsection, which shows how the
DSR requirements impose that we only consider the positive-$\lambda$ case,
illustrates even more clearly that, even restricting our attention
to the simple one-particle sector, the requirements for the
mathematical consistency of a $\kappa$-Poincar\'{e} Hopf
algebra are in general not sufficient to obtain structures which
are compatible with the DSR requirements.

Let us see this by considering the $\kappa$-Poincar\'{e} Hopf
algebra that comes closest to playing a role in the DSR scenario
of Subsection~2.2. This is the $\kappa$-Poincar\'{e} Hopf
algebra with commutation relations $\kappa \sim 1/\lambda$\footnote{Here the
careful reader will recognize the structure of the Majid-Ruegg
bicrossproduct basis~\cite{majrue}.}
\begin{eqnarray}
\left[ P_{\mu},P_{\nu} \right]&=&0\nonumber\\
\left[ M_j,M_k \right] &=&i\varepsilon_{jkl}M_l\;\;
\left[ N_j,M_k \right]=
i\varepsilon_{jkl}N_l \;\;
\left[ N_j,N_k \right] =-i\varepsilon_{jkl}M_l \nonumber \\
\left[ M_j,P_0 \right] &=&0\;\;\;
[M_j,P_k]=i\epsilon_{jkl}P_l \nonumber \\
\left[ N_j,P_0 \right]&=&i P_j \nonumber \\
\left[ N_j,P_k \right]&=&i
\left[\left( \kappa \frac{1 - e^{2 P_0 /\kappa}}{2}
+\frac{\vec{P}^2}{2 \kappa} \right) \delta_{jk}
-\frac{P_j P_k}{\kappa} \right]
~.
\label{algerelMR}
\end{eqnarray}
While a Lie algebra is fully specified by the commutation relations,
a Hopf algebra also involves a ``co-algebra sector", which
encodes the rules for the action of generators on tensor products
of one-particle states:
\begin{eqnarray}
\Delta(P_0)&=&P_0\otimes 1+1\otimes P_0\;\;\;\Delta(P_i)
=P_i \otimes 1
+e^{- P_0 / \kappa} \otimes P_j \nonumber \\
\Delta(M_i)&=&M_i\otimes 1+1\otimes M_i \nonumber \\
\Delta(N_j)&=&N_j\otimes 1+e^{- P_0 / \kappa}\otimes N_j
+\frac{1}{\kappa}\epsilon_{jkl} P_k \otimes M_l
\label{coalgerelMR}
\end{eqnarray}

The connection between these algebraic relations and
the DSR scenario of Subsection~2.2 is most easily seen by using
the fact that a Casimir of (\ref{algerelMR}) is
\begin{equation}
2 \kappa^2 \cosh (P_0 / \kappa) - \vec{P}^2 e^{ P_0 / \kappa}
~.
\label{casimirMR}
\end{equation}
The correspondence with the DSR dispersion relation (\ref{dispKpoin})
is seen upon $\kappa \rightarrow 1 / \lambda$, $P_0  \rightarrow E$,
$ \vec{P} \rightarrow  \vec{p}$.

However, the Hopf-algebra requirements do not specify in any
way the sign of $\kappa$ whereas instead the DSR requirements
impose, as we showed in the previous subsection, $\lambda \ge 0$.
Both for $\kappa >0$ and for $\kappa <0$
the relations (\ref{algerelMR}),(\ref{coalgerelMR})
satisfy the Hopf algebra criteria~\cite{majrue,kpoinap}.
This ``sign issue" is actually not surprising: the pathology
for $\lambda < 0$ which we encountered in the previous subsection emerged
at the level of analysis of finite boost transformations,
while none of the Hopf algebra requirements pertains to (or can be related
with) the concept of a finite symmetry transformation\footnote{Even in
the well-understood subject of Lie-algebra symmetries the algebra
relations (the commutation relations) are not directly related
with the description of finite symmetry transformations (they are however
used directly in the description of infinitesimal symmetry
transformations).}.
In other  ``$\kappa$-Poincar\'{e} bases" the Lorentz sector of the commutation
relations is actually modified, with commutators among Lorentz (boost/rotation)
generators that depend also on the translation generators.
While a deformed (nonlinear) action of boosts is a natural possibility
for a DSR scenario, it appears~\cite{gacdsr,dsrnext,leedsr}
that in order to have
a consistent DSR scheme the deformation must be such that
the commutators among rotation/boost generators still close the
Lorentz algebra.
For the cases in which the ``$\kappa$-Poincar\'{e} basis"
involves a modification of the Lorentz-sector commutators
it appears~\cite{gacdsr,dsrnext,leedsr} that there is no hope of
ending up with a DSR scenario.
The result of the previous subsection shows that even
in some cases (at least in the $\kappa < 0$ case of the basis
we are considering) in which the Lorentz-sector commutators
are not modified a ``$\kappa$-Poincar\'{e} basis"  leads
to an inconsistency with the DSR requirements.

Of course, the reverse is also true: some structures which are acceptable
from the DSR perspective are not consistent with
the $\kappa$-Poincar\'{e}. A well known example of this possibility
is the fact that, as stressed in Ref.~\cite{lukieNEWdsr},
the energy-momentum-conservation law (\ref{conservnewe})-(\ref{conservnewp}),
which is perfectly ok from a DSR perspective,
is not compatible with the $\kappa$-Poincar\'{e} mathematics
(in particular, the law (\ref{conservnewe})-(\ref{conservnewp})
is symmetric under exchange of the incoming or outgoing particles,
whereas $\kappa$-Poincar\'{e} mathematics requires~\cite{lukieNEWdsr}
that this law should not be symmetric\footnote{Note however that,
while in the relevant $\kappa$-Poincar\'{e} literature~\cite{lukieNEWdsr}
it is firmly argued that the coproduct should not be symmetric,
there are some works on Hopf algebras which, at least in the
case of 1+1-dimensional systems, considered the possibility of
a symmetrized coproduct~\cite{bone}.}.

In summary, while indeed certain specific
technical aspects of some DSR proposals
and some  $\kappa$-Poincar\'{e} proposals are closely related
(most notably a possible modification of the dispersion relation),
the conditions for a DSR structure are in general clearly
different from the conditions for a $\kappa$-Poincar\'{e} structure.
The DSR idea concerns a modification of the laws of transformation
that connect arbitrarily different inertial observers, in such
a way that the large-velocity scale $c$ and the small-length scale $\lambda$
(large energy-momentum scale $1 / \lambda$)
both acquire the status of relativistic invariants that characterize
these transformations.
The  $\kappa$-Poincar\'{e}  idea concerns the possibility of finding
a Hopf algebra which involves a scale $\kappa$ and in
the $\kappa \rightarrow \infty$ limit reduces to the Poincar\'{e}
Lie algebra.
As seen through the example of the law of energy-momentum conservation
(\ref{conservnewe})-(\ref{conservnewp}), some structures that
are perfectly acceptable from the DSR perspective may not satisfy
some key $\kappa$-Poincar\'{e} requirements.
And analogously, as seen in the simplest way in the case of
the ``sign of $\kappa$" issue, certain possibilities which are
fully compatible with the $\kappa$-Poincar\'{e} requirements
are inadmissable from a DSR perspective.

\section{Photon stability}

Although from a conceptual perspective they are extremely significant,
Planck-scale departures from Lorentz symmetry lead to negligible
small new effects in ordinary situations, when the energies of
the particles involved are much smaller than the Planck scale.
As it will also emerge from the phenomenological analysis reported
in this paper, in order to reach an observable large level the
relevant Planck-scale effects must be ``amplified"~\cite{polonpap}
by the presence,
in the data being considered, of a large (cosmological) distance
and/or the Planck-scale suppression must be softened by the
presence of some particles of ultra-high energies, as sometimes
happens in the context of certain observations in astrophysics.
We will assume in our presentation that the reader
is familiar with the analyses
(see, {\it e.g.}, Refs.~\cite{polonpap,qgpSS,qgpDA}
and references therein)
which showed that in controlled on-Earth laboratory setups
the new effects introduced by
Planck-scale departures from Lorentz symmetry are always
negligibly small.
However, it is also emerging that in certain contexts of interest
in astrophysics one can test the different scenarios for
the fate of Lorentz symmetry at the Planck scale.
The key opportunities come from observations which provide
indirect evidence of the stability of the photon,
observations which involve synchrotron radiation,
observations which are sentivite to the precise form of
the threshold conditions for particle production in collision processes,
and observations that are sensitive to a possible wavelength dependence
of the speed of photons.
We shall consider all of these four opportunities, starting
in this section with photon stability.

\subsection{Brief review of the broken-Lorentz-symmetry result}
It has been recently realized~\cite{jaco,jaconature,gacpion,seth,orfeupion}
that when Lorentz symmetry is broken at the Planck scale
there can be significant implications for certain decay processes.
One of the decay processes whose analysis leads to a striking result
is photon decay into an electron-positron pair: $\gamma \rightarrow e^+ e^-$.
Let us analyze this process using the dispersion relation (\ref{displead})
and unmodified energy-momentum conservation, {\it i.e.} in the
context of the broken-Lorentz-symmetry scenario of Subsection~2.1.
One easily finds a relation between
the energy $E_\gamma$ of the incoming photon, the opening angle $\theta$
between the outgoing electron-positron pair, and the energy $E_+$ of
the outgoing positron (of course the energy of the outgoing electron
is simply given by $E_\gamma - E_+$).
For the region of phase space with $m_e \ll E_\gamma \ll E_p$
one easily finds
\begin{eqnarray}
\cos(\theta) &\! \simeq \!& \frac{E_+ (E_\gamma -E_+) + m_e^2
- \eta  E_\gamma E_+ (E_\gamma -E_+)/E_p}{ E_+ (E_\gamma -E_+)} ~,
\label{gammathresh}
\end{eqnarray}
where $m_e$ is the electron mass.

The fact that for $\eta = 0$ Eq.~(\ref{gammathresh}) would
require $cos(\theta) > 1$ reflects the fact that if Lorentz symmetry
is preserved the process $\gamma \rightarrow e^+ e^-$ is kinematically
forbidden. For $\eta < 0$ the process is still forbidden, but for
negative positive $\eta$ high-energy photons can decay
into an electron-positron pair. In fact,
for $E_\gamma \gg (m_e^2 E_p/|\eta |)^{1/3}$
one finds that $\cos(\theta) < 1$, {\it i.e.} there is a physical
phase space available for the decay.

The energy scale $(m_e^2 E_p)^{1/3} \sim 10^{13} eV $ is not
very high. The fact that certain observations in astrophysics
allow us to establish~\cite{jaco,seth,jaconature}
that photon of energies up to $\sim 10^{14}eV$
are not unstable implies that the case of positive $\eta$
is severely constrained.
Since the positive-$\eta$ case, for this broken-Lorentz-symmetry
scenario, is already disfavoured conceptually (see Subsection~3.1),
we shall, without further hesitation, assume that $\eta \leq 0$
in the broken-Lorentz-symmetry
scenario of Subsection~2.1.

\subsection{No photon instability in DSR}
The fact that one finds that a certain particle decay
can occur only at energies higher than a certain minimum
decay energy ($E_\gamma \gg (m_e^2 E_p/|\eta |)^{1/3}$)
is of course a manifestation of the break down of Lorentz symmetry.
A given photon will have high energy for some inertial observers
and low energy for other inertial observers.
And of course it is not possible\footnote{We are of course
assuming the ``objectivity of particle-production processes"~\cite{gacdsr}.
If according to one observer the ``in state" (a time ``$- \infty$")
is a photon and the ``out state" (a time ``$+ \infty$")
is composed of an electron-positron pair, then all other observers
must agree.}
 to assume that the decay would
be allowed according to some observers and disalloed according to others.
So clearly such a picture requires a preferred frame: the energy of
the particle should be measured in the preferred frame and the
decay is allowed if the energy of the particle in the preferred frame
exceeds a certain given value.

For the particle-decay picture of the previous subsection the
existence of a preferred class of inertial observers is therefore
a prerequisite. And even without any calculations we can conclude
that there is no such mechanism in a consistent DSR scenario,
where the presence of preferred frames is excluded by construction.

For completeness we can verify explicitly that
the process $\gamma \rightarrow e^+ e^-$ is not allowed in
the DSR scenario of Subsection~2.2.
We must simply combine the dispersion relation (\ref{dispKpoin})
(which coincides with the dispersion relation (\ref{displead})
of the broken-symmetry scenario upon the identification $\eta = \lambda E_p$)
with the DSR-deformed energy-momentum conservation law
(see (\ref{conservnewe})-(\ref{conservnewp}))
which in this case takes the form
\begin{eqnarray}
E_\gamma \simeq E_+ + E_- - \lambda \vec{p}_+ {\cdot} \vec{p}_- ~,~~~
 \vec{p}_\gamma \simeq \vec{p}_+ + \vec{p}_- - \lambda E_+ \vec{p}_-
 - \lambda E_- \vec{p}_+ ~.
\label{consGAMMAdecay}
\end{eqnarray}
From this, considering again the region of phase space
with $m_e \ll E_\gamma \ll E_p \sim 1 / \lambda$,
one easily finds that the relation between $E_\gamma$,
the opening angle $\theta$, and $E_+$ must take the form
\begin{eqnarray}
\cos(\theta) &\! \simeq \!& \frac{2 E_+ (E_\gamma -E_+) + 2 m_e^2
+ \lambda  E_\gamma E_+ (E_\gamma -E_+)}{2 E_+ (E_\gamma -E_+)
+ \lambda  E_\gamma E_+ (E_\gamma -E_+)} ~,
\label{gammathreshDSR}
\end{eqnarray}
Evidently in this DSR case, no matter how large is $E_\gamma$,
one inevitably finds\footnote{We notice that this result
formally applies both to the positive-$\lambda$ and the negative-$\lambda$
cases. But of course, since (as shown in Subsection 3.2) the consistency
of the DSR scheme requires $\lambda > 0$, the case of true interest
is positive-$\lambda$.}
that the process would
require $\cos(\theta) > 1$, {\it i.e.} there is no physical phase
space available for the process $\gamma \rightarrow e^+ e^-$.

We conclude that the fact that certain observations in astrophysics
allow us to establish
that photons of energies up to $\sim 10^{14}eV$
are not unstable does not introduce any constraints on the parameters
of a DSR scenario. It led to an important constraint for the
broken-Lorentz-symmetry scenario of Subsection 2.1, but it is
completely unconsequential for the DSR scenario of
Subsection 2.2, even though the two scenarios adopt the same
modification of the dispersion relation.

\section{Synchrotron radiation}
\subsection{Brief review of the broken-Lorentz-symmetry analysis}
As observed recently in Ref.~\cite{jaconature}
(and observed already earlier in Ref.~\cite{gonzasync})
in the mechanism that leads to the production of
synchrotron radiation a key role is played by the
special-relativistic velocity law $v = dE/dp \simeq 1 - m^2 /(2 E^2)$.
Making the natural assumption that the relation $v = dE/dp$ still holds
at the Planck scale the modified dispersion relation (\ref{displead})
replaces the special-relativistic velocity law, with
\begin{equation}
v \simeq 1 - \frac{m^2}{2 E^2} + \eta \frac{E}{E_p}
~.
\label{velLIV}
\end{equation}
Assuming that all other aspects of the analysis of synchrotron
radiation remain unmodified at the Planck scale (an assumption which
of course must be subject to careful further scrutiny), one
is led~\cite{jaconature} to the conclusion that, if $\eta < 0$, the
energy/wavelength dependence of the Planck-scale term in (\ref{velLIV})
can severely affect the value of the cutoff energy for
synchrotron radiation~\cite{jackson}.
In fact, for $\eta < 0$, an electron on, say, a circular trajectory
(which therefore could emit synchrotron radiation)
cannot have a speed that exceeds the maximum
value
\begin{equation}
v^{max}_{e} \simeq 1
- \frac{3}{2} \left( |\eta | \frac{m_e}{E_p} \right)^{2/3}
~,
\label{velLIVmax}
\end{equation}
whereas in special relativity $v^{max}_{e} = 1$ (although values
of $v_e$ that are close to $1$ require a very large electron energy).

This may be used to argue that for negative $\eta$
the cutoff energy for synchrotron radiation
should be lower than it appears to be suggested by certain
observations of the Crab nebula~\cite{jaconature}.
In the near future, when the relevant observations of the Crab nebula
will be more precise and better understood and the assumptions
that are used in the analysis of Ref.~\cite{jaconature}
will be more carefully examined, could allow to set a very stringent
limit ($|\eta| \ll 1$) on the negative values of $\eta$ for the
broken-Lorentz symmetry scenario of Subsection~2.1.
Therefore, since positive values of $\eta$ for the
broken-Lorentz symmetry scenario are already ruled out (see Subsections~3.1
and 4.1), these planned studies of the Crab nebula and of synchrotron
radiation have the possibility of ruling out completely\footnote{As
mentioned, it is however possible to contemplate ``nonuniversal"
modifications of the dispersion relation, in which case
some of these points are significantly modified.
In particular, it was argued in Ref.~\cite{emnsync}
that a specific picture of spacetime foam~\cite{emnstrings}
suggests that $\eta = 0$ for the electron and $\eta \sim 1$
for photons. If $\eta = 0$ for electrons then there is
no modification of the velocity law for electrons.}
the scenario of Subsection~2.1.

\subsection{Implications of DSR for synchrotron radiation are negligible}
Assuming again
that the relation $v = dE/dp$ still holds
at the Planck scale\footnote{We must stress however that in the DSR framework
the validity of the relation $v = dE/dp$ might depend on the relevant spacetime
picture. Because of the nature of the phenomenological analyses here of interest
we could here focus on the energy-momentum sector but of course a full DSR
theory must include a description of the spacetime sector.
In particular, for one of the spacetimes which are being considered as
candidate DSR-compatible spacetimes, the so-called $\kappa$-Minkowski
noncommutative spacetime, the validity of the relation $v = dE/dp$
is still a subject of an active
debate~\cite{frangianlu,Kosinski:2002gu,Mignemi:2003ab,Daszkiewicz:2003yr,jurekREV}.
We restrict here our attention to the relation $v = dE/dp$
because we are not trying to examine a variety of DSR scenarios,
but we are rather focusing on the DSR scenario of Subsection 2.2,
and on an analysis of that DSR scenario which makes it as
similar as possible
to the broken-Lorentz-symmetry scenario of Subsection 2.1
(in order to expose some aspects of phenomenology which are genuinely
sensitive to the difference between breaking and deforming Lorentz symmetry).}
from the DSR modified dispersion relation (\ref{dispKpoin})
one obtains the velocity law
\begin{equation}
v \simeq 1 - \frac{m^2}{2 E^2} + \lambda E
~.
\label{velDSR}
\end{equation}
For $\lambda < 0$ one would also in this case find that for
a particle of mass $m$ there is an $m$-dependent
(and smaller than 1) maximum value of the velocity.
However, we have seen in Subsection~3.2 that negative values
of $\lambda$ are not compatible with the logical structure
of the DSR framework, since they require the existence
of a preferred class of inertial frames.
And for positive $\lambda$, since there is no maximum value
of the speed of electrons, the formula (\ref{velDSR})
does not suggest any modification of the cutoff energy
for synchrotron radiation.

It is perhaps useful to summarize our findings before the additional
phenomenological analyses of the following Sections~6 and 7.
Our first level of analysis, in Section~3, focused
on the logical structure of the scenarios, and we found that
in the DSR scenario of Subsection 2.2 it was necessary to
impose $\lambda >0$, while for the broken-Lorentz-symmetry scenario
of Subsection 2.1 there was no absolute requirement
on the sign of $\eta$, although $\eta \leq 0$ appeared to be preferable.
We then, in Section~4, considered on photon-stability issues, and we
found that some relevant data rule out $\eta > 0$
in broken-Lorentz-symmetry scenario of Subsection 2.1,
while they have no significant implication for the analysis of
the DSR scenario of Subsection 2.2.
So at the end of Section~4 we had ruled out $\eta > 0$
and $\lambda < 0$, but negative values of $\eta$ and positive
value of $\lambda$ were still completely unconstrained.
Then, here in Section~5, we focused on synchrotron-radiation
issues, and we found that some relevant data (when better understood)
provide an opportunity to rule out the residual possibility
for the broken-Lorentz-symmetry scenario of Subsection 2.1,
{\it i.e.} the case of negative $\eta$, while the same
synchrotron-radiation data are not significant for the
positive-$\lambda$ DSR scenario of Subsection 2.2.
In the remainder of this paper we will focus on positive values
of $\lambda$ for the DSR scenario,
and we will continue to consider negative
values of $\eta$ for the broken-Lorentz-symmetry scenario.
Indeed our results so far leave still completely
uncostrained the positive values of $\lambda$.
For $\eta$ we are in a situation in which positive values
are clearly excluded (through the photon-stability analysis)
but negative values are still allowed (pending a more careful
examination of the synchrotron-radiation analysis
of Ref.~\cite{jaconature}).

\section{Threshold anomalies}
\subsection{Brief review of the broken-Lorentz-symmetry result}
Another opportunity to investigate Planck-scale departures
from Lorentz symmetry is provided by certain types of energy
thresholds for particle-production processes that are relevant in
astrophysics.
A simple way to see this is found in the analysis
of collisions between
a soft photon of energy $\epsilon$
and a high-energy photon of energy $E$ that creates an electron-positron
pair: $\gamma \gamma \rightarrow e^+ e^-$.
For given soft-photon energy $\epsilon$,
the process is allowed only if $E$ is greater than a certain
threshold energy $E_{th}$ which depends on $\epsilon$ and $m_e^2$.
In the broken-Lorentz-symmetry scenario of Subsection~2.1
this threshold energy can be  evaluated combining
the dispersion relation (\ref{displead}) with ordinary
energy-momentum conservation. One easily obtains
(assuming $\epsilon \ll m_e \ll E_{th} \ll E_p$)
\begin{equation}
E_{th} \epsilon + \eta \frac{E_{th}^3}{8 E_p}= m_e^2
~.
\label{thrTRE}
\end{equation}
The special-relativistic result $E_{th} = m_e^2 /\epsilon$
corresponds of course to the $\eta \rightarrow 0$ limit
of (\ref{thrTRE}).
For $|\eta | \sim 1$ the Planck-scale correction can be
safely neglected as long as $\epsilon /E_{th} > E_{th}/E_p$.
But eventually, for sufficiently small values of $\epsilon$ and
correspondingly large values of $E_{th}$, the
Planck-scale correction cannot be ignored.
This occurs for $\epsilon < (m_e^4 /E_p)^{1/3}$
(when, correspondingly $E_{th} > (m_e^2 E_p)^{1/3}$).
For $\epsilon \sim 0.01 eV$
the modification of the threshold is already significant,
and this is relevant for the observation
of multi-$TeV$ photons from certain Blazars~\cite{aus,gactp}.
Blazar photons travel to us from very far
and they travel in an environment populated by soft photons,
some with energies (roughly $\epsilon \sim 0.01 eV$)
suitable for acting as targets
for the disappearance of
the hard photon into an electron-positron pair.
According to the standard ($\eta = 0$) threshold formula,
Blazar photons of energies in the neighborhood of $10 TeV$
should be already significantly absorbed before reaching our
Earth telescopes.
If $\eta$ is negative (the only case we are still considering, since
positive $\eta$ is safely ruled out) and of order $1$ one finds
from (\ref{thrTRE}) a significantly higher value of the threshold
(varying $\eta$ between, say, 0.1 and 10 one finds that the threshold
can be shifted up to $20 TeV$ and higher).
Therefore the availability of good-quality data on the observed
spectrum of these Blazar photons, as we should surely have at least once
the space telescope GLAST~\cite{glast} starts operating in 2006,
could be used to set stringent constraints on negative $\eta$.

A similar argument can be applied to cosmic rays.
The fact that ultra-high-energy cosmic-ray protons travel toward
the Earth in an environment populated by CMBR photons
leads to the expectation of a GZK~\cite{gzk}
cutoff on the observed spectrum.
In the standard special-relativistic framework the GZK cutoff
is around $5 10^{19} eV$, whereas in the broken-Lorentz-symmetry
scenario with dispersion relation (\ref{displead})
one finds a modification of the threshold~\cite{kifu,ita,gactp}
analogous to (\ref{thrTRE}).
And again for negative $\eta$ of order $1$ one finds
a  value of the threshold that is  significantly higher than
the GZK value (comfortably above $10^21 eV$).
Good-quality data on the observed
spectrum of cosmic rays at energies of order $10^{20}$ or $10^{21} eV$
would therefore allow to test a key prediction of the negative-$\eta$
case.
At present the situation could be described as encouraging,
since some observations reported by
AGASA~\cite{agasa}, arguably the best cosmic-ray observatory
presently taking data, provide encouragement for the idea of
a cutoff scale significantly higher than the GZK prediction.
But it is necessary to proceed with caution: there are other
(non-Planck-scale-related) possible
descriptions~\cite{stecksync,bererev}
of the AGASA anomaly, and also the experimental situation
requires further scrutiny (since the findings of the
HIRES observatory~\cite{hires,bacCR} are inconsistent with ones of AGASA).
Important elements for this analysis should become available in
the coming months, when the Pierre Augerre cosmic-ray observatory
should announce its first data.

\subsection{Threshold anomalies are typically small in DSR}
While, as shown in the previous subsection, ``threshold anomalies"
can be observably large in the broken-Lorentz-symmetry scenario
of Subsection 2.1, in DSR the threshold anomalies are typically
negligibly small. In particular, in the DSR scenario of Subsection 2.2
one finds negligibly small threshold anomalies.

One can easily verify this by analyzing again the
process $\gamma \gamma \rightarrow e^+ e^-$.
In the DSR scenario of Subsection 2.2 one should analyze the
kinematics of this process using
again the dispersion relation (\ref{dispKpoin})
(which is essentially the same as the dispersion relation (\ref{displead})),
but taking into account~\cite{gacdsr,sethdsr}
the DSR-deformed energy-momentum conservation law
(see (\ref{conservnewe})-(\ref{conservnewp}))
which in this case takes the form
\begin{eqnarray}
E + \epsilon - \lambda \vec{P} {\cdot} \vec{p} \simeq E_+
+ E_- - \lambda \vec{p}_+ {\cdot} \vec{p}_- ~,~~~
 \vec{P} +  \vec{p}
 - \lambda E \vec{p}
 - \lambda \epsilon \vec{P}  \simeq \vec{p}_+ + \vec{p}_-
 - \lambda E_+ \vec{p}_-
 - \lambda E_- \vec{p}_+ ~
\label{consgammagamma}
\end{eqnarray}
where we denoted with $\vec{P}$ the momentum of the photon of energy $E$
and we denoted with  $\vec{p}$ the momentum of the photon
of energy $\epsilon$.

The presence of correction terms both in the dispersion relation
and in the energy-momentum-conservation law (with coefficients fixed
by the requirement of equivalence of inertial frames)
leads to rather large cancellations in the threshold formula.
Assuming again that $\epsilon \ll m_e \ll E_{th} \ll E_p$ one ends up
finding
\begin{equation}
E_{th} \simeq \frac{m_e^2}{\epsilon }
~,
\label{thrTREbis}
\end{equation}
{\it i.e.} (for $\epsilon \ll m_e \ll E_{th} \ll E_p$) one ends up
with the same result as in the special-relativistic case.
If, rather than working within the approximations allowed
by the hierarchy $\epsilon \ll m_e \ll E_{th} \ll E_p$,
one considers the exact DSR threshold formula, one finds
a result which is actually
different from the special-relativistic one. There
are ``threshold anomalies" in the DSR scenario of Subsection 2.2.
But they are very small, well below an observable level,
when $\epsilon \ll m_e \ll E_{th} \ll E_p$, which corresponds to the
only real opportunity for testing such threshold anomalies.

An analogous result holds for the cosmic-ray threshold.
The calculation is somewhat more tedious but we find that the end
result is again affected by large cancellations in the approximations
that are natural when the energy of the cosmic ray is much higher that
the proton mass and much smaller than the Planck scale (and also
using the fact that the energies of CMBR photons are much smaller
than the proton mass).
For $\epsilon_{CMBR} \ll m_{proton} \ll E_{cosmic-ray} \ll E_p$
one finds only a minute difference (well below the level observable
at cosmic-ray observatories) between the special-relativistic
threshold and the DSR threshold.

We conclude that planned studies of ``threshold anomalies"
at gamma-ray and cosmic-ray observatories will not allow to discriminate
between special relativity and the DSR scenario of Subsection 2.2.

We must stress here that there is no in-principle obstruction
for observably large threshold anomalies in DSR. It just happens to
be difficult to construct a DSR scenario
with observably large threshold anomalies.
We have verified (and the reader can easily verify) even considering
the possibility of changing
rather arbitrarily the dispersion relation, if done consistently
with the DSR requirements ({\it i.e.} introducing a corresponding
modification of the energy-momentum conservation law)
one nearly {\underline{inevitably}} finds a final result
in which the threshold anomalies are negligibly small.
But if one really desperately wants observably large threshold
anomalies, even at the cost of a completely {\it ad hoc} choice
of the DSR dispersion relation, a suitable DSR scenario can be
found, and indeed such a scheme was proposed in Ref.~\cite{dsrgzk}.
This is the scenario which we already mentioned in discussing
Eq.~(\ref{dispgood}) (and the ``creative" presence
of $\tanh$ functions in Eq.~(\ref{dispgood}) reflects
the  {\it ad hoc} choices that guided the construction of that scenario).

So, in summary, the threshold anomalies of the DSR scenario
of Subsection 2.2 (with dispersion relation (\ref{dispKpoin}))
are negligibly small, and the threshold anomalies are also negligibly
small in other DSR scenarios, but with some {\it ad hoc} choices
in introducing the elements of a DSR scenario one can manage to
produce sizeable threshold anomalies.

This status of threshold anomalies in DSR is rather different
from the results of our analysis of photon stability in DSR.
We have verified explicitly that in the most studied DSR scenario,
the one we discussed in Subsection 2.2, photon decay
into an electron-positron pair is not allowed and there are no
sizeable threshold anomalies. But for photon decay we observed that
one can exclude it in any, however structured, DSR scenario because
it would conflict with the required equivalence of inertial observers,
whereas the presence of sizeable threshold anomalies is not in
conflict with any general requirement of the DSR framework (they simply
turns out to be absent in the DSR scenario of Subsection 2.2,
and we found that some rather {\it ad hoc} choices must be made
in constructing a DSR scenario if one has the objective of introducing
sizeable threshold anomalies).

\section{Time-of-travel analyses}
In Section~5 we already noticed that both in the broken-Lorentz-symmetry
scenario of Subsection~2.1 and in the DSR scenario of Subsection~2.2
one finds a Planck-scale contribution to the energy dependence
of the speed of particles.
It is convenient to introduce a unified notation here
for the relevant velocity laws (for $m < E \ll E_p \sim 1/ \lambda$):
\begin{equation}
v \simeq 1 - \frac{m^2}{2 E^2} + \eta_* \frac{E}{E_p}
~,
\label{velLIVbis}
\end{equation}
where $\eta_* = \eta \leq 0$ for the
broken-Lorentz-symmetry scenario
and $\eta_* = \lambda E_p \geq 0$ for the DSR scenario.

In Section~5 we were focusing on synchrotron radiation, where a
significant effect might be expected if $\eta_* < 0$
(with $| \eta_* | \sim 1$), so the analysis turned out to be only relevant
for the broken-Lorentz-symmetry scenario.
There is of course another, more direct, way to investigate
the possibility (\ref{velLIVbis}):
whereas in ordinary special relativity two photons ($m=0$)
with different energies
emitted simultaneously would reach simultaneously a far-away detector,
those two photons should reach the detector at different times
according to (\ref{velLIVbis}).

This type of effect emerging from
an energy dependence of the speed of photons
can be significant~\cite{grbgac,billetal}
in the analysis of short-duration gamma-ray bursts that reach
us from cosmological distances.
For a gamma-ray burst it is not uncommon\footnote{Up to 1997 the
distances from the gamma-ray bursters to the Earth were not
established experimentally.
By a suitable analysis of the
gamma-ray-burst ``afterglow"~\cite{beppoSAX},
it is now possible to establish the distance
from the gamma-ray bursters to the Earth for a significant portion of
all detected bursts. $10^{10}$ light years ($\sim 10^{17} s$) is
not uncommon.} to find a time travelled
before reaching our Earth detectors of order $T \sim 10^{17} s$.
Microbursts within a burst can have very short duration,
as short as $10^{-3} s$ (or even $10^{-4} s$), and this means that the photons
that compose such a microburst are all emitted at the same time,
up to an uncertainty of $10^{-3} s$.
Some of the photons in these bursts
have energies that entend at least up to the $GeV$ range.
For two photons with energy difference of order $\Delta E \sim 1 GeV$
an $\eta_* \Delta E/E_p$ speed difference over a time of travel
of $10^{17} s$
would lead to a difference in times of arrival of
order $\Delta t \sim \eta_* T \Delta E/E_p \sim 10^{-2} s$, which
is significant (the time-of-arrival differences would be larger than
the time-of-emission differences within a single microburst).

Such a Planck-scale-induced time-of-arrival difference
could be revealed~\cite{grbgac,billetal}
upon comparison of the structure of the gamma-ray-burst signal
in different energy channels.
The next generation of gamma-ray telescopes,
such as GLAST~\cite{glast},
will exploit this idea to search for
energy dependence of the speed of photons.

An even higher sensitivity to a possible
energy dependence of the speed of photons could be achieved
by exploiting the fact that, according to
current models~\cite{grbNEUTRINOnew},
gamma-ray bursters should also emit a substantial amount of
high-energy neutrinos.
Some neutrino observatories should soon observe neutrinos with energies
between $10^{14}$ and $10^{19}$ $eV$, and one could, for example, compare
the times of arrival of these neutrinos emitted by
gamma-ray bursters to the corresponding times of arrival of
low-energy photons. But a robust estimate
of the sensitivity achievable following
this strategy will require an improved understanding
of gamma-ray bursters, good enough establish whether there are typical
at-the-source time delays (there could be a systematic tendency of
gamma-ray bursters to emit high-energy
neutrinos with, say, a certain delay with
respect to microburst of photons).

In any case, this type of time-of-arrival analyses (both the ones
that use exclusively photons and the ones that might be able to
use also neutrinos) is clearly important both for
the broken-Lorentz-symmetry scenario and for the DSR scenario,
since in both scenarios one expects an energy-dependent
speed of photons.
Our analysis (particularly the results reported in Subsection~3.2)
provides a new tool for these experimental studies.
In fact, we can now establish that data
in favour of $\eta_* < 0$ would not only signal a departure
from Lorentz symmetry but would also suggest that there
is a break down of Lorentz symmetry at the Planck scale
(since $\eta_* < 0$ is not allowed in the DSR scenario).
And {\it vice versa} data
in favour of $\eta_* > 0$ would not only signal a departure
from Lorentz symmetry but would also suggest that there
is a DSR-deformation of Lorentz symmetry at the Planck scale
(since the possibility of  $\eta_* > 0$ in the broken-Lorentz-symmetry
scenario is disfavoured by the phenomenological analysis of
Section~4).

\section{Closing remarks}
Our analysis shows that
the possibility of a Planck-scale deformation of Lorentz
symmetry, in the sense of DSR,
leads to a phenomenology which in many ways is significantly
different from the more studied corresponding
scenario with broken Lorentz symmetry.
In some contexts in which the broken-Lorentz-symmetry scenario
could lead to
observably-large departures from the predicitions of the
ordinary Lorentz-symmetry case, as in the observations relevant
for photon stability, synchrotron radiation and anomalous thresholds,
the DSR scenario is instead basically
indistinguishable from ordinary special relativity.
These differences are due to the fact that in DSR
one still has a 6-parameter rotation/boost symmetry group,
and this introduces various limitations to the new effects that
may arise. For example, as we discussed in Section~4 (on the basis
of the points already raised in Ref.~\cite{gacdsr}),
just like in special relativity also in DSR
it is not possible for a particle which is stable at low
energies to become unstable above some high-energy scale.
The same features of DSR kinematics (energy-momentum conservation)
which enforce this principle end up playing a role also
in the analysis of the possibility of anomalous thresholds (Section~6).
A sizeable modification of the threshold relations
for particle-physics reactions is not {\it a priori} excluded
by the DSR requirements, but since the analysis of particle decays
and the analysis of reaction thresholds share many computational aspects
it ends up being the case that the same structures
that protect DSR from unadmissable decay properties also render
typically small the modifications of the reaction thresholds.

A DSR scenario is instead not much different from
a broken-Lorentz-symmetry scenario for what concerns the type
of time-of arrival analyses discussed in Section~7.

This leads us to a rather sharp characterization of the possibility
of departures from Lorentz symmetry in which the dispersion relation
involves a leading-order term that goes linearly with the Planck scale,
as the ones of Subsections~2.1 and 2.2.
Ordinary special relativity of course predicts that no ``anomaly"
should be found in the observations considered in Sections~4, 5, 6 and 7.
The DSR scenario of Subsection 2.2 predicts that there should be
no anomalies for what concerns the observations considered
in Sections~4, 5 and 6 but there should be a trace of the Planck-scale
effects in the time-of-arrival analyses of Section~7.
The broken-Lorentz-symmetry scenrio of Subsection~2.1 definitely
predicts observably-large anomalies in all of the types
of observations considered in Sections~4, 6 and 7 and possibly
also the ones considered in Section~5 (pending further scrutiny
of some of the assumtpions in the analysis
of the implications of broken Lorentz symmetry for
synchrotron radiation reported in Ref.~\cite{jaconature}).

Clearly the investigation this possibility
of departures from Lorentz symmetry in which the dispersion relation
involves a leading-order term that goes linearly with the Planck scale
is within the reach of the gamma-ray and cosmic-ray observatories
that should start taking data in the next 4 or 5 years.
On the basis of the analysis we reported here, one can be confident
that with these observatories the possibility of such
Planck-scale-linear modifications will be fully explored.
We will either be able to rule it out completely on the basis of
actual data or find definite evidence of such new effects. And if the
data do provide support for the new Planck-scale
effects their analysis, on the basis of the characterization
we provided here, will also allow us to distinguish between
the DSR scenario and the broken-Lorentz-symmetry scenario.

Of course, once the possibility of effects
that go linearly with the Planck scale has been fully explored
(and, as one tends to expect, possibly ruled out),
the next item on the agenda of investigation of the fate
of Lorentz symmetry at the Planck scale will concern the
possibility of effects
that go quadratically with the Planck scale. Since these
quadratically-suppressed effects are obviously much smaller than
the linear ones, their investigation with actual data is somewhat further
in the future. The fact that instead data relevant for the linear case
is soon becoming available motivated us to focus on that possibility,
but we did mention, although somewhat parenthetically, that
some opportunities for testing the quadratic case might not be too far
in the future.
From that perspective a key role should be played by
cosmic-ray observations and the powerful tool
that could be provided by high-energy-neutrino observatories~\cite{grf03}.

On the technical side our result showing that $\lambda$ must be positive
in the DSR scenario of Subsection~2.2 could have several applications,
and already proved to be rather poweful (together with the presence
of the 6-parameter group of rotation/boost symmetries)
for our phenomenological analysis. This result for the sign of $\lambda$
should also contribute to a deeper understanding of the DSR conceptual
structure, and in particular, as we stressed, should allow to resolve
completely the frequent confusion between the DSR proposal and
the mathematics of $\kappa$-Poincar\'{e} Hopf algebras.

\section*{Acknowledgments}
The work of JKG was supported in part by the KBN
grant 5PO3B05620. We also acknowledge INFN support for visits within the
collaboration. GAC and AP are greatful to A.~Degasperis
for feed-back on these results.

\baselineskip 12pt plus .5pt minus .5pt

\vfil

\end{document}